How to use 2D gel electrophoresis in plant proteomics

Thierry Rabilloud

CNRS UMR 5249, Chemistry and Biology of Metals, Grenoble
Université Joseph Fourier, Grenoble
iRTSV/LCBM, CEA Grenoble
17 rue des martyrs, F-38054 Grenoble Cedex 9, France
mail: thierry.rabilloud@cea.fr

Abstract

Two-dimensional electrophoresis has nurtured the birth of proteomics. It is however no longer the exclusive setup used in proteomics, with the development of shotgun proteomics techniques that appear more fancy and fashionable nowadays. Nevertheless, 2D gel-based proteomics still has valuable features, and sometimes unique ones, which make it often an attractive choice when a proteomics strategy must be selected. These features are detailed in this chapter, as is the rationale for selecting or not 2D gel-based proteomics as a proteomic strategy.

1. Introduction

At the beginning of the 21st century, it is fashionable to describe 2D gel-based proteomics as an outdated and poorly efficient technique. This feeling is mainly due to the fascination of many scientists in proteomics for big lists, and in this case the well-documented undersampling of 2D gel-based proteomics is dearly resented, as well as its almost complete inability to analyze transmembrane proteins [1, 2], despite of lot of effort devoted to this particular issue (e.g. in [3, 4] for plant samples). This undersampling has been documented mostly on animal samples [5, 6], but it is obvious that it will take place on plant samples as well.

In fact, all proteomic techniques undersample complex samples by a ratio of 1:10 (1000 proteins analyzed on 2D gels among the 10,000 present in a cell, 20,000 peptides analyzed by shotgun proteomics among the 200,000 derived from a cell sample). However, as 2D gel-based proteomics is the only setup that analyses complete proteins, the consequences of undersampling are clear and heavy. If a protein is not seen, it is lost for ever. In addition, the undersampling rules in 2D gel-based proteomics are well known. Hydrophobic proteins, high molecular weight proteins and rare proteins do not show up.

In shotgun techniques, the undersampling rules are also known. However, they apply to peptides, and how they translate in terms of protein identification and even more in terms of protein quantification is complex, and linked to the complexity of the protein inference problem [7-9] .

On top of this strong undersampling problem comes the often unrecognized fact that 2D gel-based proteomics is a low yield process [10, 11] and the fact that plant samples are among the most difficult samples to prepare for an analysis by 2D gels [12], due to the variety and amounts of interfering, non-protein compounds present in many plant cell types and tissues.

All these negative sides explain the trend from 2D gel-based proteomics to shotgun proteomics, as exemplified by the study of cadmium response in plant. Gel-based in 2006 [13], shotgun-based in 2011 [14].

However, 2D gel-based proteomics is still very popular in plant proteomics, as shown by recent publications in the field (e.g. [15, 16], but among more than 150 publications in 2011). There are several built-in reasons for this sustained interest in this technique, which are named reproducibility, robustness, efficiency, ability to analyze intact proteins, adequacy to study post-translational modifications, easy interface with powerful biochemical techniques. These different points will be developed below.

2. Reproducibility and robustness

It is not an overstatement to write that 2D gel-based proteomics is still, and by far, the most reproducible and robust proteomic setup [17]. A further proof of this statement lies in the much higher expectations from proteomic journals, in terms of sample numbers and experimental power, applied to 2D gel-based proteomics compared to shotgun proteomics [18, 19]. Still today, what is common practice in 2D gel-based proteomics, i.e. quantitative comparison of two or more biological conditions with several individual biological replicates for each condition, is still rarely found in publications using shotgun proteomics.

This reproducibility is further testified by the interlaboratory reproducibility of 2D maps [20, 21], while recent test of shotgun techniques on the same sample (yeast) revealed a much higher standard deviation between laboratories [22, 23]

Ironically enough, the high robustness of 2D gel-based proteomics comes in part from the terrible reproducibility of isoelectric focusing in its early days. Put bluntly, only the gels run simultaneously were comparable. This led the proteomic pioneers to develop devices to run several gels in parallel [24, [5]. When immobilized pH gradients were finally introduced in 2D gel electrophoresis with solid protocols [26], the reproducibility and robustness of the technique made a quantum leap that is still enjoyed today. Moreover, this parallelization of the technique is also a very strong advantage when multiple comparisons, and not only binary ones, must be performed [27].

## 3. Efficiency of 2D gel-based proteomics

Nowadays, Edman sequencing for protein identification has been almost completely replaced by mass spectrometry-based protein identification, which is much more sensitive and much faster. However, mass spectrometers are expensive and delicate instruments, and the machine time is often considered as precious.

In this frame, shotgun proteomic techniques often represent a waste of machine time. In many comparative experiments, which represent most of the proteomic experiments, most of the proteins do not change between the various conditions tested. Thus, the mass spectrometer is analyzing again and again a very high number of peptides from proteins that have no interest in the biological question studied.

Oppositely, 2D gel-based proteomics makes an optimal use of mass spectrometer time, because most of what is carried out within the mass spectrometer in shotgun protemics is carried out upfront at the 2D gel stage in 2D gel-based proteomics. In fact, the modern gel staining techniques [28-30], allow to perform a sensitive and linear detection of the protein spots on the gels. Then, image analysis [31, 32] coupled with statistical analysis [33], allow to perform a quantitative analysis of all protein spots and to determine which spots are worth identifying. These spots are then the only ones that need to be processed by the mass spectrometer.

When financial resources are scarce, which is often the case in academic research, this allows to build a "hub and spokes" model, where a central hub hosting the mass spectrometry platform is able to serve very efficiently several biology-oriented laboratories, each using 2D gel electrophoresis and image analysis to select the relevant proteins within their respective research projects.

With plant proteomics usually less funded than medicine-oriented proteomics, and this situation even harder for the proteomics of traditional crops, this efficiency factor cannot be neglected in the research landscape.

## 4. Interface with biochemical techniques

Within this section, the focus will be shifted from classical, quantitative proteomics,

where the name of the game is to find differences in expression on total extracts, to more focused proteomics experiments where the interest is centered on subclasses of proteins. In many cases, the protein subclasses are defined by a common biochemical feature, and quite often this biochemical feature is a post-translational modification, for which an affinity reagent exist.

When such an affinity reagent exists, two types of strategies can be devised to perform proteomic studies targeted for the proteins recognized by the affinity reagent.

In the first strategy, the affinity reagent is used as a pre-proteomic preparative tool, aiming at selecting the proteins or peptides of interest in the complex starting extract. The selected proteins/peptides are then analyzed by a classical proteomic setup.

This strategy is the only one that can be used with shotgun proteomics, but it can also be used with 2D gel-based proteomics as well.

In the second strategy, the affinity reagent is used as an analytical reagent, after the separation stage in the proteomic setup, in order to detect selectively the proteins of interest. This strategy works extremely well with 2D gel-based proteomics, thanks to the easy interface provided by the blotting process between 2D gels and affinity reagents.

Thus, the final performance of the whole process will largely depend on the efficiency of the affinity reagent under the preparative and analytical setups, and it must be kept in mind that for several reasons, the affinity reagent is always used in a large excess over the biological extract.

Then two major cases can be distinguished:

i) the affinity reagent is not a protein

typical examples of that configuration are represented by boronic acid for sugars and either IMAC or metal oxides for phosphate. These non-proteinaceous reagents are best used in the preparative mode, and work wonderfully with shotgun proteomics [34].

ii) the affinity reagent is a protein

In this case, the suitability of the reagent as a preparative tool will depend not only on its selectivity, but also on the existence of mild elution conditions that are able to elute all the analytes without eluting out the affinity reagent.

Lectins represent such a happy case, and have used as a preparative reagent in 2D gel-based proteomics [35] and in shotgun proteomics [36-38].

They have also been used as a post-2D gel selective detection reagent [39]

Conversely, antibodies represent a case where elution without polluting the extract with antibodies is very difficult. They are therefore much better used as analytical reagents on blots rather than preparative reagents. Thus, 2D gel-based proteomics is a very efficient strategy to use in conjunction with antibodies, and this strategy has been used to identify several post-translational modifications, such as citrullinated proteins[40], carbonylated proteins [41], proteins containing nitrotyrosine [42, 43] or hydroxynonenal adducts [44, 45].

Although very successful, this strategy has a major caveat, i.e. the fact that a minor and modified spot detected with the antibody may comigrate with a major, unrelated spot which will be the one identified by the downstream mass spectrometry process, resulting in a major mistake. The severity of this problem increases with spot crowding and thus with the complexity of the biological extract. There is however an easy counter, at least for acidic and neutral proteins, which is the use of narrow range pH gradients that greatly decrease the probability of comigration. Such narrow pH range gels have been used under a variety of circumstances [46-49], and have proven very efficient, especially when used in conjunction with prefractionation techniques [50].

5. Analysis of intact proteins and of post-translational modifications

The more proteomics progresses, the more it stresses the importance of post-translational modifications in biology (e.g. in [51, 52] ). However, not all post-translational modifications can be easily studied through the use of an affinity reagent, and some, as methylations, are not easily detected except on abundant proteins [53].

For the study of these difficult, and sometimes unknown, modifications, 2D gel-based proteomics is a platform of choice, either through selective and transient labelling [54], or by using the fact that many modifications induce a pH shift and thus a separation of the modified protein from the bulk of the unmodified one. This can be used in the field of protein cleavage [55, 56] , but also as a preparative tool to isolate the modified spot, thereby leading to an easier identification of the modifications. Such an approach has been used for the identification of thiol oxidations [57-59].

More generally, the separating ability of 2D gels, coupled with their ability to be used as a micropreparative tool [60], now allow to set very comprehensive maps of protein modifications [61, 62]. In this area, the use of ultra narrow pH gradient [47], with their almost infinite resolving power [63, 64] should be extremely useful, by providing a very thorough deconvolution of the protein spots into individual protein species.

6. Conclusions

Nowadays, 2D gel-based proteomics is usually selected on the basis of its robustness and relatively low cost, among the proteomic techniques. It can be reasonably anticipated, however, that the uses directed toward the identification of post-translational modifications, and especially the determination of the real combination of modifications present on a single protein species, will develop considerably in the future.